\documentclass{PoS}
\usepackage{IEEEtrantools}

\title{Lattice calculation of the HVP contribution to the anomalous magnetic moment of muon}

\ShortTitle{Lattice calculation to the HVP contribution to g-2}

\author{\speaker{Bipasha Chakraborty}\\SUPA, School of Physics and Astronomy, University of Glasgow, Glasgow G12 8QQ, UK\\
        E-mail: \email{b.chakraborty.1@research.gla.ac.uk}}
\author{Christine Davies\\SUPA, School of Physics and Astronomy, University of Glasgow, Glasgow G12 8QQ, UK}
\author{Pedro Gon\c{c}alves de Oliveira\\SUPA, School of Physics and Astronomy, University of Glasgow, Glasgow G12 8QQ, UK}
\author{Jonna Koponen\\SUPA, School of Physics and Astronomy, University of Glasgow, Glasgow G12 8QQ, UK}
\author{G. Peter Lepage\\Laboratory of Elementary Particle Physics, Cornell University, Ithaca, NY 14853, USA}
\author{HPQCD Collaboration}
                              
\abstract{We report our (HPQCD) progress on the calculation of the Hadronic Vacuum Polarisation contribution to the anomalous magnetic moment of muon. In this article we discuss the calculations for the light (up/down) quark connected contribution using our method described in Phys.Rev. D89 (2014) 11, 114501 and give an estimate for the disconnected contribution. Our calculation has been carried out on MILC Collaboration's $n_f$ = 2+1+1 HISQ ensembles at multiple values of the lattice spacing, multiple volumes and multiple light sea quark masses (including physical pion mass configurations).}

\FullConference{The 33rd International Symposium on Lattice Field Theory\\
                 14-18 July, 2015\\
                 Kobe International Conference Center, Kobe, Japan}

\begin{document}

\section{Motivation} 
\label{sec:intro2}

\begin{figure}
\centering
\includegraphics[width=0.28\textwidth]{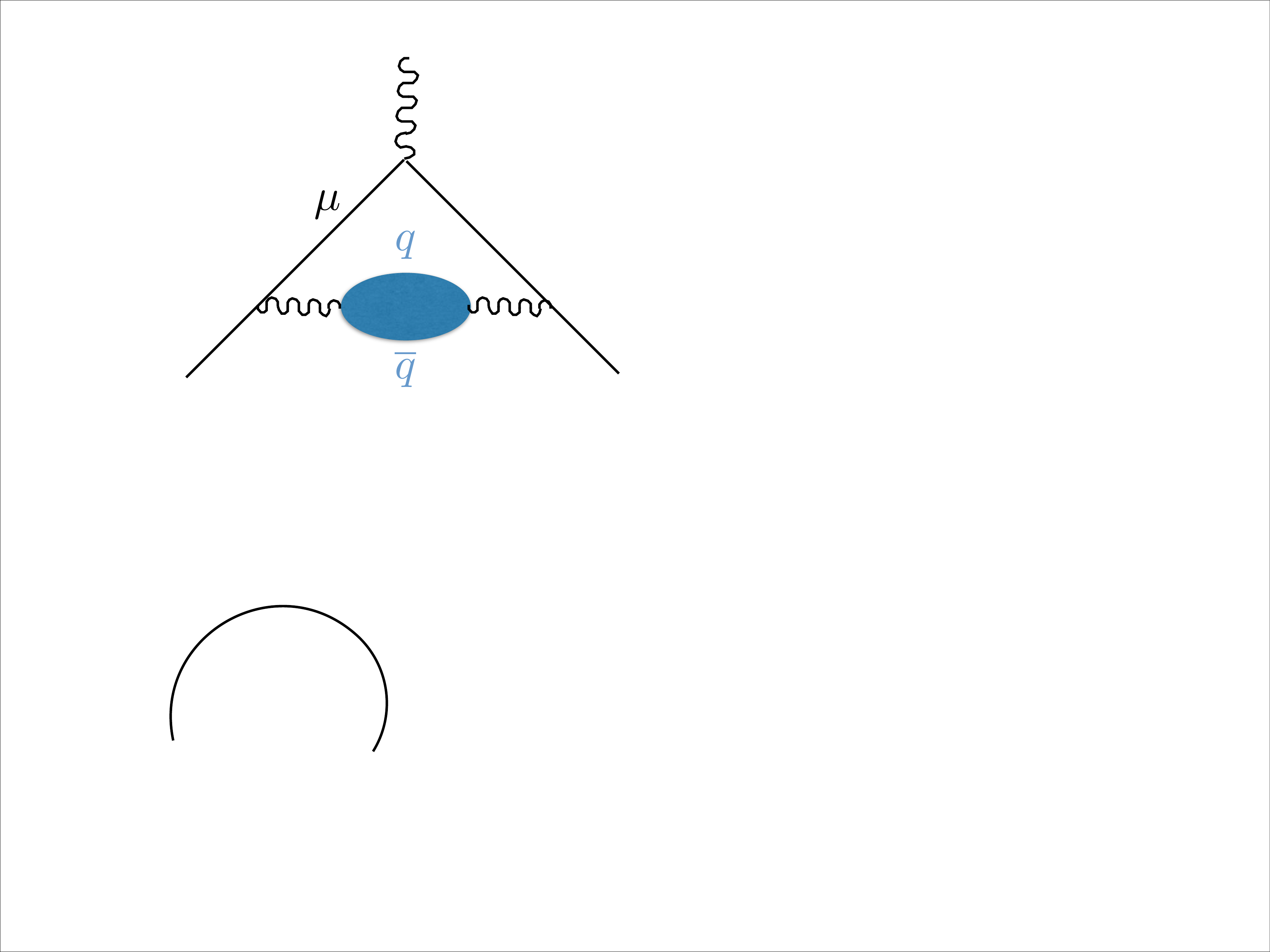}
\caption{The hadronic vacuum polarization contribution to the muon anomalous magnetic moment is represented as a shaded blob inserted into the photon propagator (represented by a wavy line) that corrects the point-like photon-muon coupling at the top of the diagram.}
\label{fig:hvp}
\end{figure}
The anomalous magnetic moment of the muon ($a_\mu$), defined as the fractional difference of its gyromagnetic ratio from the naive value of 2, ($a_{\mu}=(g-2)/2$), gives one of the most stringent tests of the Standard Model. This quantity is measured in experiment with impressive precision, 0.54 ppm~\cite{Bennett:2006fi} and shows a tantalizing discrepancy of 3$\sigma$ with the Standard Model expectation: $a_\mu^{exp}-a_\mu^{SM}=25(9) \times 10^{-10}$ ~\cite{Aoyama:2012wk, Hagiwara:2011af,rafael1972, Davier}. With the forthcoming experiments at Fermilab and J-PARC aiming to reduce the experimental uncertainty by a factor of 4, it is now vital to achieve a comparable precision from theory. The current theoretical uncertainty is dominated by that from the theoretical calculation of the lowest order ``hadronic vacuum polarisation (HVP)'' contribution, $a_{\mu}^{HVP, LO}$. This contribution is currently determined most accurately using the dispersion relation and the experimental results on $e^+e^- \rightarrow$ hadrons or from $\tau$ decay to be of size $700 \times 10^{-10}$ with $\sim$1$\%$ error~\cite{Hagiwara:2011af, Davier}. Our goal is to achieve an uncertainty of less than 1$\%$ in $a_{\mu}^{HVP,LO}$ using a first principle lattice QCD calculation. We have developed a simple lattice QCD method~\cite{PhysRevD.89.114501} for calculating $a_{\mu}^{HVP,LO}$ which improves significantly on previous calculations and using that method we have achieved 1$\%$ uncertainty in the strange quark-line connected contributions to the HVP~\cite{PhysRevD.89.114501}. In this article we report on our progress with the calculation of the light quark connected piece using the same method and of the disconnected piece of the HVP contribution to $a_{\mu}^{HVP,LO}$ and quote our preliminary result for the total $a_{\mu}^{HVP,LO}$.

\section{Simulation Details}
We calculate the light (up/down) quark propagators using the Highly Improved Staggered Quark (HISQ)~\cite{HISQ_PRD} discretisation on HISQ gauge configurations generated by the MILC collaboration~\cite{Bazavov:2010ru, Bazavov:2012uw} with light, strange and charm quarks in the sea. We have used three lattice ensembles with lattice spacings a $\approx$ 0.15\,fm (very coarse), 0.12\,fm (coarse) and 0.09\,fm (fine), determined~\cite{fkpi} using the Wilson flow parameter $w_0$~\cite{Borsanyi:2012zs}. At each lattice spacing we have three values of the average $u/d$ quark mass: $m_s/5, m_s/10, m_s/27.5$ (physical). On $m_l = m_s/10$ and a $\sim$ 0.12\,fm we have three different volumes corresponding to a lattice length in units of the $\pi$ meson mass of $M_{\pi}L=$ 3.2, 4.3 and 5.4 to test for finite volume effects. 

The light quark propagators are combined into a correlator with a local vector current at either end. The end point is summed over spatial sites on a timeslice to set the spatial momentum to zero. We use a random colour wall source created from a set of U(1) random numbers over a timeslice for improved statistics. The local current is not the conserved vector current for the HISQ quark action and must be renormalised. We have calculated the local vector current renormalisation constant ($Z_{V,\overline{s}s}$) completely non-perturbatively with 0.1$\%$ uncertainty on the finest $m_l=m_s/5$ lattices\cite{Chakraborty:2014zma} for the strange-strange currents. For the time being we are using the same renormalisation for the light-light local vector current. 


\section{Connected light correlators}


The light quark contribution in $a_\mu^{HVP,LO}$ is the most significant part, being 12 times larger than that for the strange quark, in part because of a factor of 5 from the electric charges. Though the extension of our method~\cite{PhysRevD.89.114501} to calculate $a_{\mu}^{light}$ is straightforward, poor signal-to-noise ratio in this case significantly increases the statistical uncertainties in the time moments~\cite{PhysRevD.89.114501}. We have overcome this issue by calculating the time moments from the reconstructed correlators using the best fit parameters for time slices larger than t* (instead of using the original correlators). This constrains the errors in the correlators at larger times therefore giving a much better precision in the moments.
We used a data-fit hybrid correlator as follows:
\begin{equation}
G(t) = \left\{ \,
\begin{IEEEeqnarraybox}[][c]{l?s}
\IEEEstrut
G_{data}(t) & for $t<=t*$ from Monte Carlo, \nonumber\\
G_{fit}(t) & for $t>t*$ from multi-exponential fit.\nonumber
\IEEEstrut
\end{IEEEeqnarraybox}
\right.
\label{eq:example_left_right1}
\end{equation}
for t* = 1.5fm = 6/$m_\rho$. Thus we get 70$\%$ of the result from $G_{data}$. We get the same results to within $\pm \sigma /4$ with t* = 0.75fm.
We also improved fitting uncertainties by using Gaussian smearings at source and sink and fitting a 2 x 2 matrix of correlators. 

\begin{figure}
  \centering
  \includegraphics[width=7.0cm,angle=0]{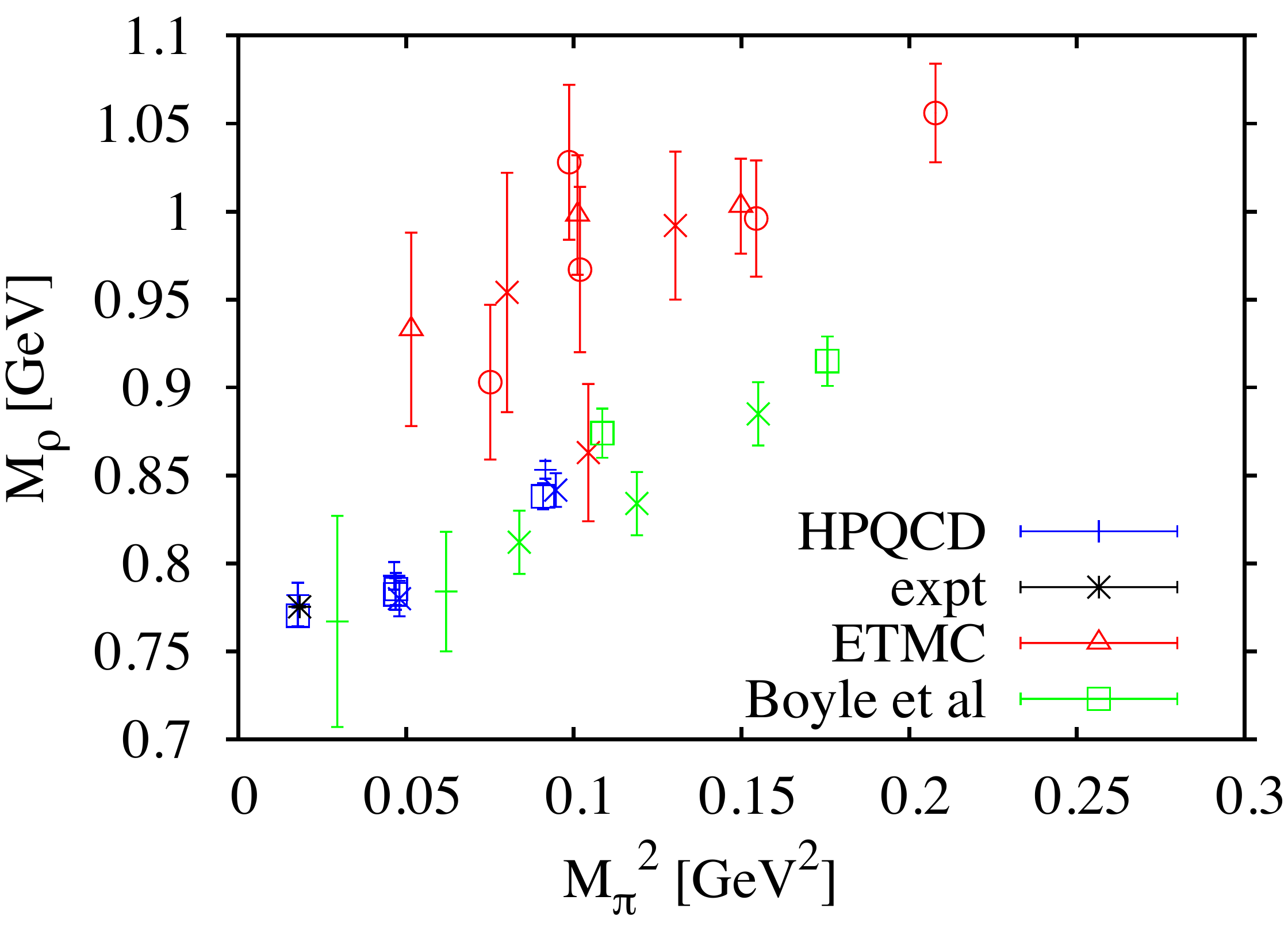}
  \includegraphics[width=7.0cm,angle=0]{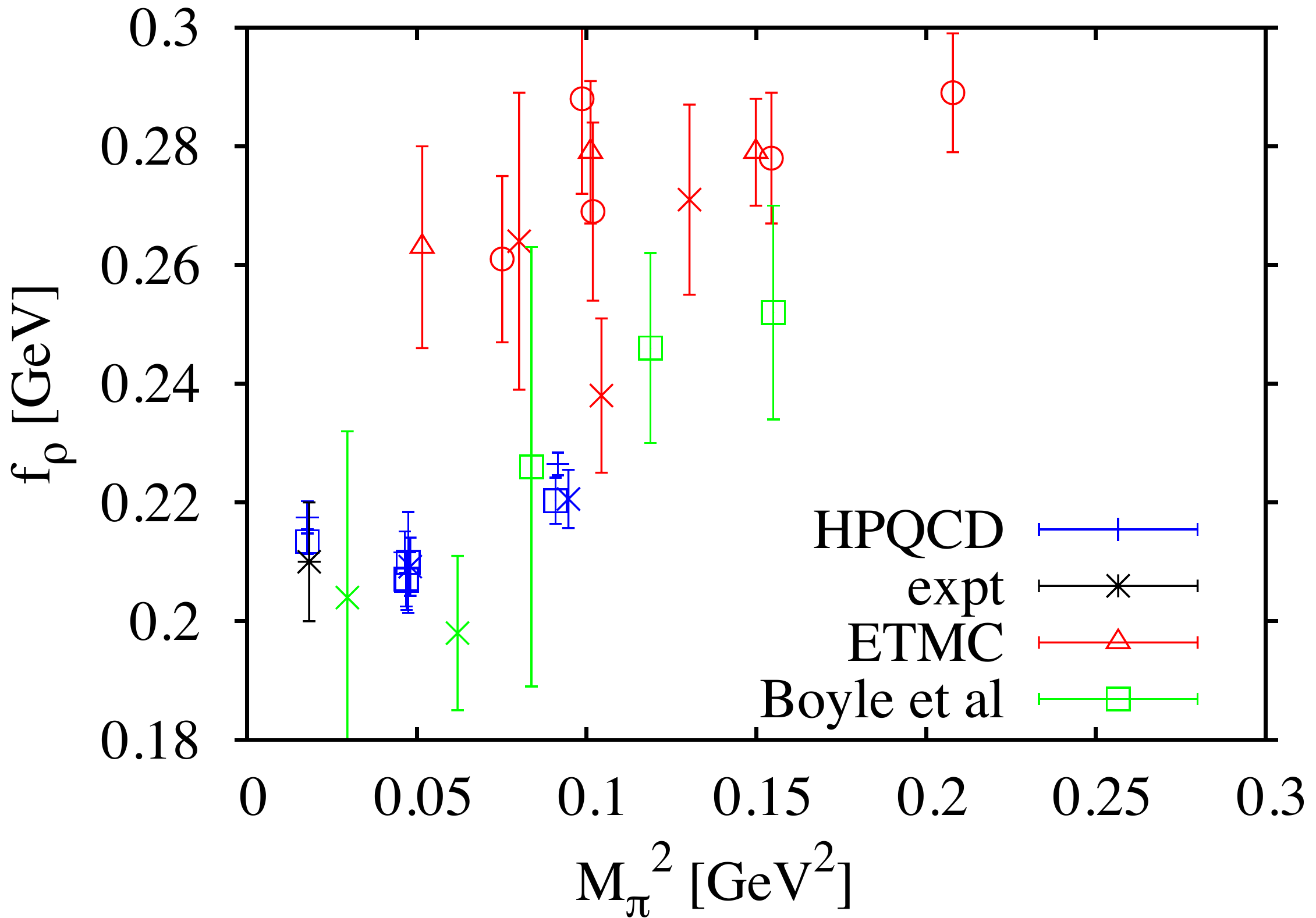}
\caption{Comparison plots of masses and decay constants of the $\rho$ meson against squared pion mass from calculations that have also been used to determine the HVP contribution to $a_\mu^{HVP}$: Results include values at multiple lattice spacings and volumes - HPQCD (a $\sim$ 0.09-0.15 fm, L $\sim$ 2.5-5.8 fm), ETMC (a $\sim$ 0.06-0.08 fm, L $\sim$ 2.5-2.9 fm), Boyle et. al. (a $\sim$ 0.09-0.14 fm, L $\sim$ 2.7-4.6 fm). Our numbers (HPQCD) in Blue, ETMC results~\cite{Burger:2013jya} in Red, results from Boyle et. al. in Green~\cite{Boyle:2011hu} and experimental results in Black}
\label{fig:mrho_frho}
\end{figure}

Our fits to the vector correlators give the decay constant and mass of the vector meson $\rho$. Our results for these are shown in Fig.\ref{fig:mrho_frho} and compared to previous lattice values. We see that our $\rho$ mass and decay constant fall towards the experimental values as the pion mass is reduced to its physical value.



Our method determines the expansion in $q^2$ of the vacuum polarisation function from the time-moments. These are dominated by the contribution from the ground state $\rho$ meson. Hence much of the light quark mass dependence comes from that of the $\rho$. We can remove this by rescaling the coefficients by appropriate powers of the $\rho$ mass~\cite{Burger:2013jya}. An important contribution which should not be rescaled in this way is that from the photon coupling to $\pi\pi$. Therefore, we first remove this $\pi\pi$ contribution on each lattice using one-loop, staggered quark, finite-volume chiral perturbation theory~\cite{Jegerlehner:2011ti}, and then restore it from one-loop continuum chiral perturbation theory, with the physical $\pi$ mass. The scaling of $\hat{\Pi}_j^{latt}$ in this way gives:


\begin{eqnarray}
\hat{\Pi}_j^{latt} \rightarrow (\hat{\Pi}_j^{latt} - \hat{\Pi}_j^{latt}(\pi\pi)) \bigg[\frac{m_\rho^{2j,latt}}{m_\rho^{2j,expt}}\bigg] + \hat{\Pi}_j^{cont}(\pi\pi)
\end{eqnarray}

\subsection{Preliminary results}

\begin{figure}[thb]
\centering
 \includegraphics[width=8.5cm,angle=0]{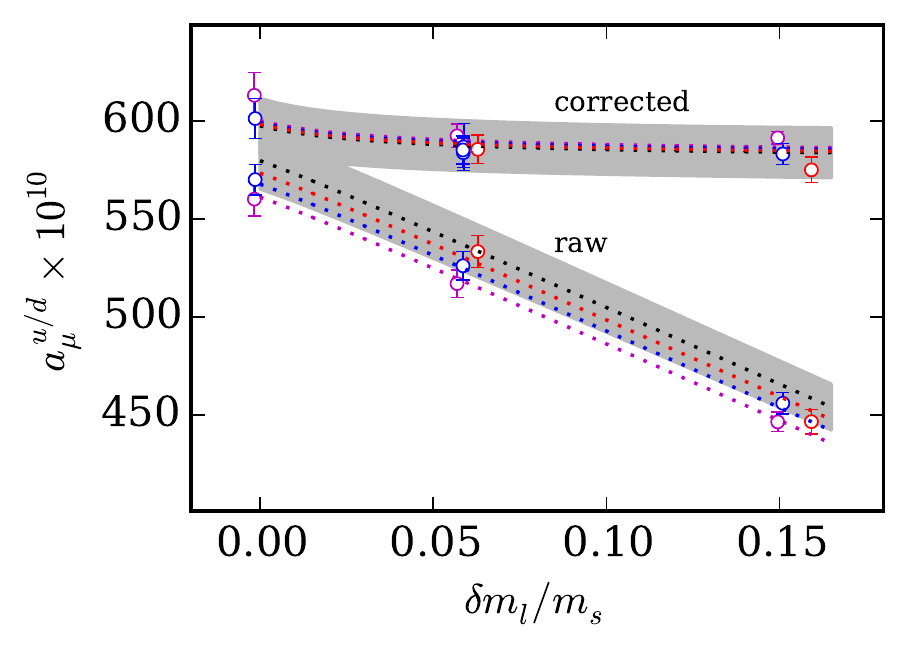}
\caption{Preliminary lattice QCD results for 
 the connected contribution to the muon anomaly $a_\mu$ from vacuum polarisation of $u/d$~quarks, both uncorrected and rescaled (corrected), for three lattice spacings, and three light-quark masses. The dashed lines are the corresponding values from the fit function, using the best-fit parameters. The gray band for the corrected results shows our final result, $598(11)\times10^{-10}$, after chiral-continuum extrapolation.}
\label{fig:l-fit}
\end{figure}

Our results for $a_{\mu}^{light}$ are shown in Fig.\ref{fig:l-fit}. The rescaled values are independent of $m_l/m_s$, $a^2$, finite volume, but the raw values also agree at the physical point. Fitting the corrected results as a function of $a^2$ and $m_{sea}$ yields a preliminary result of $a_{\mu}^{light} =$ 598(11) $\times$ 10$^{-10}$ including 1$\%$ QED and 1$\%$ isospin uncertainties (quadrature). 

\section{Disconnected correlators}

The quark-line disconnected contributions to the HVP are expected to be small since they vanish when $m_u = m_d = m_s$~\cite{Blum:2002ii}.
 On the lattice the disconnected correlators are extremely noisy.
 We have used an all-to-all propagator method with 50 stochastic noise vectors on each configuration using a one-link spatial taste-singlet vector current at both source and sink.
\begin{figure}
\centering
 \includegraphics[width=8.5cm,angle=0]{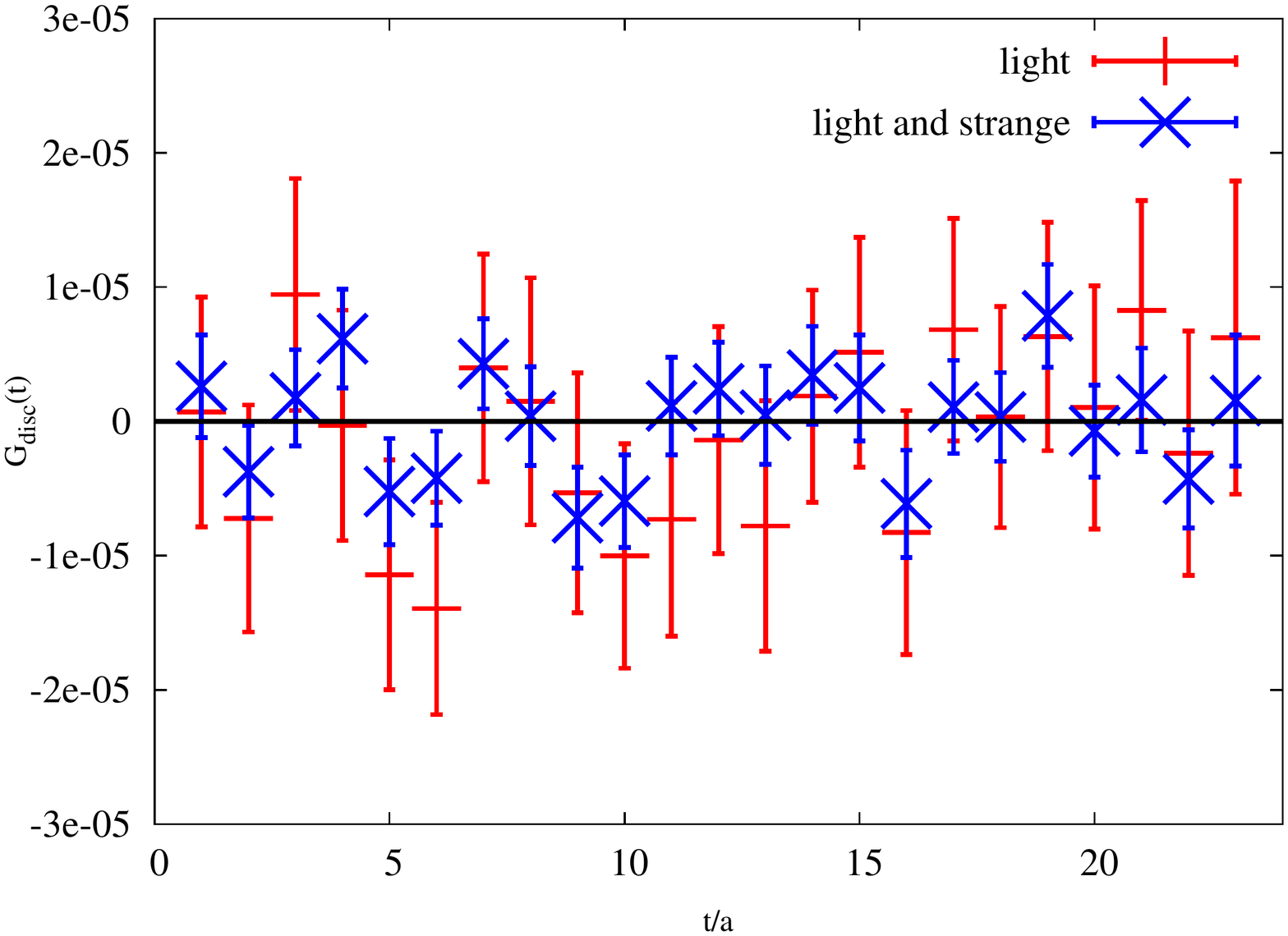}
\caption{Comparing the disconnected correlators made from light/light and from l-s/l-s currents}
\label{fig:disc}
\end{figure}
 Using the same noise for l and s currents as recommended in~\cite{Francis:2014hoa} we find a 40$\%$ reduction in uncertainty compared to the l quark case alone (see Fig.\ref{fig:disc}).

\subsection{Estimation of Disconnected contribution to g-2}

Since we have no signal we use instead an estimate for the disconnected contribution from comparing isoscalar ($\omega$) and isovector ($\rho$) correlators. At large time the disconnected light-light correlator is given by the difference between that of the $\omega$ and the $\rho$.
\begin{eqnarray}
2D_{ll,gs} = -\frac{f^2_\rho m_\rho}{2} e^{-m_\rho t} + \frac{f^2_\omega m_\omega}{2} e^{-m_\omega t}
\end{eqnarray}
The contribution to the time-moments is then readily determined. The ratio for time-moment j between $D_{ll}$ and the connected $C_{ll}$ piece is then 
\begin{eqnarray}
R_j = \frac{\Pi_{j,d}}{\Pi_{j,c}} = \frac{1}{2}\bigg[\frac{m_\rho^{2+2j}f^2_{\omega}}{m_\omega^{2+2j}f^2_{\rho}} - 1\bigg] 
\label{eq:rj}
\end{eqnarray}
The contribution to $a_{\mu}$ from $D_{ll}$ has a further factor of $1/5$ from the relative electric charges. Using experimental values: $m_\rho = 0.775$ GeV, $f_\rho = $0.21(1) GeV, $m_\omega = $0.783 GeV and $f_\omega = $0.20(1) GeV we obtain: $a_{\mu,disc}/a_{\mu,conn} \approx -1.5(1.5)\%$. Note that eq.\ref{eq:rj} trivially yields $-1/10$ for the ratio of disconnected to connected contributions to $a_{\mu}$ for the $\pi\pi$ piece, since the isoscalar channel contains no $\pi\pi$ contribution. However, the $\pi\pi$ contribution is handled using a complete calculation here, and not separated into connected and disconnected pieces.

\section{HPQCD Estimation for total $a_{\mu}^{HVP,LO}$}
\begin{figure}[b]
   \centering
   \includegraphics[width=10.5cm,angle=0]{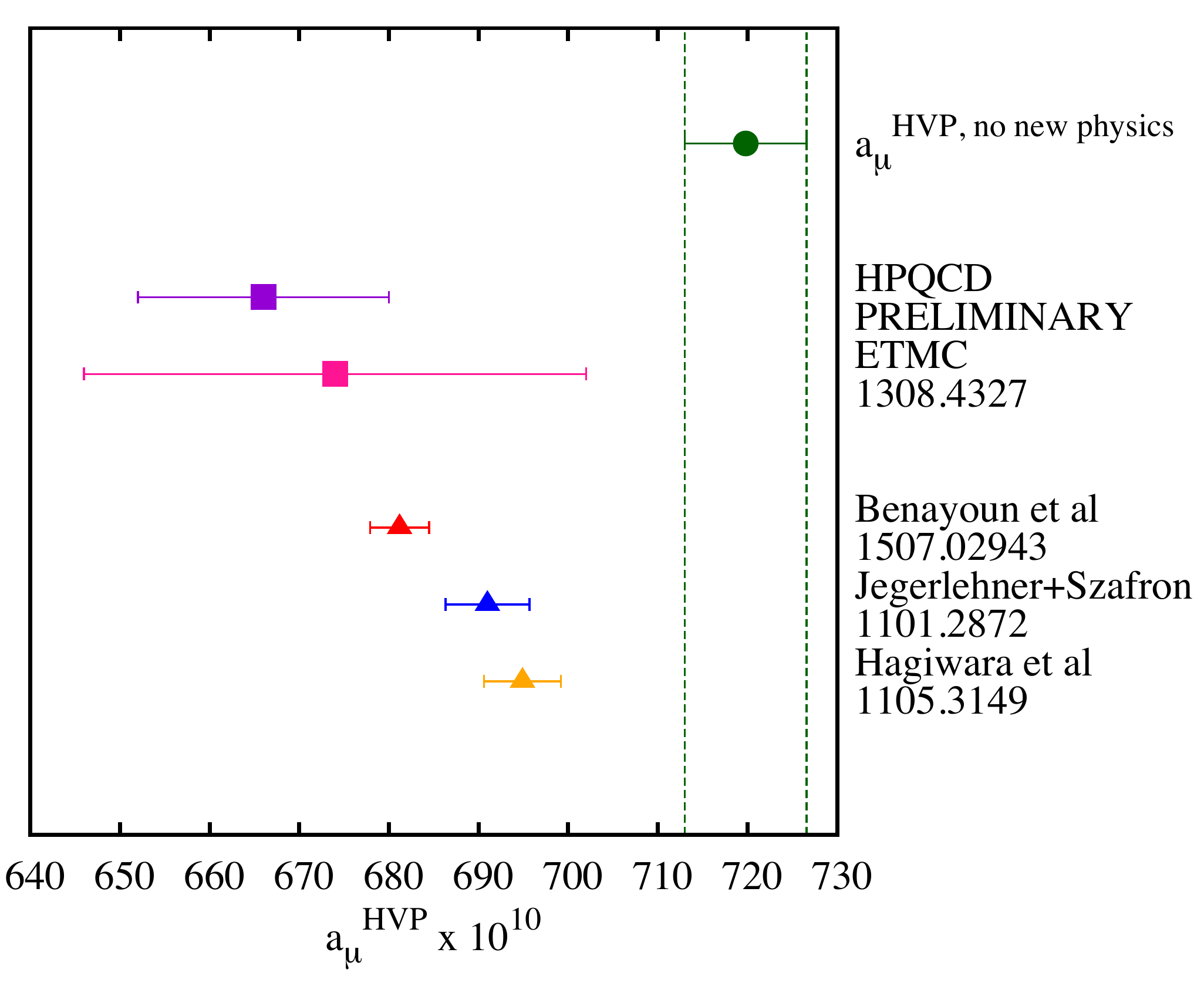}
\caption{A comparison plot showing our (HPQCD) result for $a_{\mu}^{HVP,LO}$ agrees well with other lattice (ETMC)~\cite{Burger:2013jya} and phenomelogy results~\cite{Benayoun:2015gxa}~\cite{Jegerlehner:2011ti}~\cite{Hagiwara:2011af}. Our result is 3$\sigma$ from the $a_\mu^{HVP,no new physics}$ value of 720(7) x 10$^{-10}$ obtained from the experimental result by subtracting QED, EW, HLBL and higher order HVP contributions.}
\label{fig:comp_pheno}
\end{figure}

Table \ref{tab:total} shows the contributions to HVP coming from each of the connected quark pieces and disconnected pieces. 
\begin{table}[t]
\large
\caption{This table lists our results for all the separate contributions to $a_{\mu}^{HVP,LO}$ and gives the total number (preliminary).}
\centering
\begin{tabular}{ccc}
\hline 
Contribution &  Result (x 10$^{-10}$) &\\
\hline
\hline
light, conn &  598(11) &  (preliminary including \\
            &          & 1$\%$ QED + 1$\%$ isospin effects) \\

\hline
strange, conn   & 53.4 (6)  & ~\cite{PhysRevD.89.114501}\\
\hline
charm, conn  &  14.4(4) & ~\cite{Donald:2012ga,jpsi}\\
\hline
bottom, conn & 0.27(4) & ~\cite{Colquhoun:2014ica}\\
\hline  
disconn. & 0(9) & take 1.5$\%$ as uncertainty; \\
(estimate)  &      & contribution likely to be negative\\   
\hline
\hline
Total    &  666(14) &    \\
\hline
\end{tabular}       
\label{tab:total}                                                                                                                          
\end{table}
Our preliminary estimate of the total HVP contribution to muon g-2: $a_{\mu}^{HVP,LO}$ = $666(14)\times10^{-10}$ including all systematics (with 1$\%$ QED and 1$\%$ isospin uncertainties). 
Figure \ref{fig:comp_pheno} shows that our result for $a_{\mu}^{HVP,LO}$ agrees well with other lattice (ETMC) results including u, d, s, c quarks and those using experimental cross-sections, but we have not yet achieved a level of precision comparable to that from phenomenology. 

\section{Conclusion and Ongoing Work}

The preliminary result of our calculation of connected light quark contributions to $a_{\mu}^{HVP,LO}$ using HISQ quarks gives a 1.8$\%$ uncertainty with 1$\%$ QED and 1$\%$ isospin uncertainities included. Our preliminary result for the total (u,d,s,c and b) $a_{\mu}^{HVP,LO}$ is then 666(14)$\times 10^{-10}$. Our result includes calculations with multiple lattice spacings and multiple sea quark masses including physical light quarks for the first time. 
Further work (with the MILC collaboration) will include much larger ensemble sizes and finer lattices to reduce uncertainties further.

\subsection*{\bf{Acknowledgements}}
 We are grateful to the MILC collaboration for the use of their gauge configurations. Our calculations were done on the Darwin Supercomputer as part of STFC's DiRAC facility jointly funded by STFC, BIS and the Universities of Cambridge and Glasgow. This work was funded by STFC, the Royal Society, the Wolfson Foundation and the National Science Foundation.

\bibliographystyle{h-physrev.bst}
\bibliography{g2s}{}

\begin{thebibliography}{10}

\bibitem{Bennett:2006fi}
Muon G-2 Collaboration, G.~Bennett {\em et~al.},
\newblock Phys.Rev. {\bf D73}, 072003 (2006), hep-ex/0602035.

\bibitem{Aoyama:2012wk}
T.~Aoyama, M.~Hayakawa, T.~Kinoshita, and M.~Nio,
\newblock Phys.Rev.Lett. {\bf 109}, 111808 (2012), 1205.5370.

\bibitem{Hagiwara:2011af}
K.~Hagiwara, R.~Liao, A.~D. Martin, D.~Nomura, and T.~Teubner,
\newblock J.Phys. {\bf G38}, 085003 (2011), 1105.3149.

\bibitem{rafael1972}
{B.E. Lautrup, A. Peterman and E. de Rafael, Phys. Rep. 3, N$^{\circ}$4 (1972)
  193-260}.

\bibitem{Davier}
M.~Davier, A.~Hoecker, B.~Malaescu, and Z.~Zhang,
\newblock Eur.Phys.J. {\bf C71}, 1515 (2011), 1010.4180.

\bibitem{PhysRevD.89.114501}
HPQCD Collaboration, B.~Chakraborty {\em et~al.},
\newblock Phys. Rev. D {\bf 89}, 114501 (2014).

\bibitem{HISQ_PRD}
HPQCD and UKQCD Collaborations, E.~Follana {\em et~al.},
\newblock Phys.Rev. {\bf D75}, 054502 (2007), hep-lat/0610092.

\bibitem{Bazavov:2010ru}
MILC collaboration, A.~Bazavov {\em et~al.},
\newblock Phys.Rev. {\bf D82}, 074501 (2010), 1004.0342.

\bibitem{Bazavov:2012uw}
MILC Collaboration, A.~Bazavov {\em et~al.},
\newblock Phys.Rev. {\bf D87}, 054505 (2013), 1212.4768.

\bibitem{fkpi}
HPQCD Collaboration, R.~Dowdall, C.~Davies, G.~Lepage, and C.~McNeile,
\newblock Phys.Rev. {\bf D88}, 074504 (2013), 1303.1670.

\bibitem{Borsanyi:2012zs}
S.~Borsanyi {\em et~al.},
\newblock JHEP {\bf 1209}, 010 (2012), 1203.4469.

\bibitem{Chakraborty:2014zma}
HPQCD Collaborations, B.~Chakraborty {\em et~al.},
\newblock PoS {\bf LATTICE2013}, 309 (2013), 1401.0669.

\bibitem{Burger:2013jya}
ETM, F.~Burger {\em et~al.},
\newblock JHEP {\bf 02}, 099 (2014), 1308.4327.

\bibitem{Boyle:2011hu}
P.~Boyle, L.~Del~Debbio, E.~Kerrane, and J.~Zanotti,
\newblock Phys. Rev. {\bf D85}, 074504 (2012), 1107.1497.

\bibitem{Jegerlehner:2011ti}
F.~Jegerlehner and R.~Szafron,
\newblock Eur. Phys. J. {\bf C71}, 1632 (2011), 1101.2872.

\bibitem{Blum:2002ii}
T.~Blum,
\newblock Phys.Rev.Lett. {\bf 91}, 052001 (2003), hep-lat/0212018.

\bibitem{Francis:2014hoa}
A.~Francis {\em et~al.},
\newblock PoS {\bf LATTICE2014}, 128 (2014), 1411.7592.

\bibitem{Donald:2012ga}
HPQCD Collaboration, G.~Donald {\em et~al.},
\newblock Phys.Rev. {\bf D86}, 094501 (2012), 1208.2855.

\bibitem{jpsi}
HPQCD Collaboration, C.~Davies {\em et~al.},
\newblock PoS {\bf ConfinementX}, 288 (2012), 1301.7203.

\bibitem{Colquhoun:2014ica}
HPQCD Collaboration, B.~Colquhoun, R.~J. Dowdall, C.~T.~H. Davies,
  K.~Hornbostel, and G.~P. Lepage,
\newblock Phys. Rev. {\bf D91}, 074514 (2015), 1408.5768.

\bibitem{Benayoun:2015gxa}
M.~Benayoun, P.~David, L.~DelBuono, and F.~Jegerlehner,
\newblock (2015), 1507.02943.

\end{thebibliography}

\end{document}